\documentclass[11pt]{article}
\usepackage{geometry}                
\usepackage{graphicx}
\usepackage{amsmath}
\usepackage{amssymb}
\newcommand{\povec}{V (r)}

\newcommand{\doefunds}{This work was supported in part by 
funds provided by the U.S. Department of Energy under cooperative research agreement
\#DE-FG02-05ER41360.}
\title{Challenges to Path Integral Formulations of Quantum Theories\footnote{``Path Integrals 07," Dresden, Germany, September 2007.}}
\author{R. Jackiw\\
\it \small Center for Theoretical Physics\\
\it \small MIT, Cambridge, MA 02139-4307}
\date{\small MIT/CTP-3907 }                                           
\begin{document}
\maketitle
\begin{abstract}
The functional integral has many triumphs in elucidating quantum theory.\\ But incorporating charge fractionalization into that formalism  remains a challenge.
\end{abstract}
This conference celebrates the achievements of path and functional integration in quantum physics. But it is good to remember that some of these achievements were hard to attain, because they required resolving unexpected subtleties  of the functional formalism. Two historical examples will illustrate my point, and then I shall posit a new challenge.

For the first example, we look at the description of rotationally symmetric motion in a potential  $\povec$.
The classical effective Hamiltonian reads $ H =\frac{p^2_r}{2 m} + \frac{L^2}{2m r^2} + \povec$, where the centrifugal term represents $r^2 \, \dot{\theta}^2/2m$ and vanishes for circularly symmetric motion. But in the quantum description of this motion, when it is confined to a plane, the classical centrifugal term is replaced by $\hbar^2 (M^2 - 1/4) /2 m r^2$, where $M$ is any integer. In particular a residual attraction remains for s-waves, $M=0$. This quantal attraction has the important physical consequence that in planar physics bound states exist, no matter how weakly attractive might be the potential $V$. (In this way, planar bound states follow the behavior of one dimensional bound states, rather than three dimensional, which require a minimum strength to achieve binding.)

The functional integral involves integration over c-number functions. A direct change of variables from Cartesian to circular coordinates reproduces the classical centrifugal barrier, but misses the residual, attractive potential $-\hbar^2/8 m r^2$. This is particularly vexing, since coordinate changes are point canonical transformations, which are allowed in quantum mechanics.

The resolution of this problem was given by Edwards and Gulyaev \cite{sfed1964}. But it required returning to the discretized formulation of the path integral and realizing that the angular step $\Delta \theta$, is not of order of the temporal step $\Delta\tau$, but rather $0 (\sqrt{\Delta\tau})$. [This issue reappeared in the collective coordinate quantization of solitons. Investigations based on functional integrals missed $0 (\hbar^2)$ terms that were found in an operator approach \cite{jlger1976}.]

For my second example, I turn to the anomaly phenomenon: the classical action can possess an invariance against a transformation that is not an invariance after quantization. Where is this effect in the functional integral, which involves ``functional integration" over the exponentiated classical action? The answer, which was found by Fujikawa (after anomalies were discovered by conventional methods), located the effect in the functional measure. Evidently in the anomalous situation it is not invariant; a fact that is established after the measure is discretized \cite{Fujikawa:2004cx}.

In both instances we see that the validity of formal changes of variables in a functional integral must be assessed by a return to the discretized functional sum, in fact, by a return to conventional quantum theory.

Now I shall present another peculiar quantum effect, which as far as I know has not had a functional integral description, even though a conventional quantal argument establishes  it rather easily. I have in mind the phenomenon of fermion charge fractionalization in the presence of a topological defect \cite{jrebbi1976}. 

It is well known and also guaranteed by various index theorems, that the Dirac equation in the presence of a topological defect (kink in one spatial dimension, vortex in two, monopole in three etc.; we call these ``solitons") possesses an isolated, normalized zero-energy, mid-gap bound state. Of course the Dirac equation also possesses positive energy solutions and negative energy solutions. The conventional instruction, given by Dirac, is to define the vacuum by filling the negative energy states and leaving the positive states empty; with this definition the vacuum charge vanishes. But what should one do with the mid-gap, zero-energy bound states, if it is present? Dirac is silent on this; he did not know about mid-gap bound states. The answer is that there is double degeneracy in energy, since filling the mid-gap state costs no energy.  Moreover, the empty mid-gap state carries fermion charge $-1/2$, and the filled one, $+ 1/2$. Remarkably, this effect has been observed in polyacetylene --- a 1-dimensional lineal material --- and it has been proposed for 2-dimensional  planar graphene. (Note: the Dirac equation is relevant to these condensed matter materials not because of relativistic considerations; rather a well defined linearization of the energy dispersion near the Fermi surface gives rise to a linear matrix equation, which is of the Dirac type.)

Fermion number fractionalization is established by  the following argument. Consider the Dirac Hamiltonian $H(\varphi)$ in a background field $\varphi$, which can be topologically trivial or non-trivial. In the trivial case $\varphi$ takes a homogenous value $\varphi_v$, which gives rise to a mass gap in the Dirac spectrum, between the positive and negative continuum energy eigenstates, which we call ``vacuum states" and denote them by $\psi_E$.
\begin{equation}
H (\varphi_v)\, \psi_E = E\, \psi_E \qquad E \gtrless 0
\end{equation}
With the topologically non trivial background  $\varphi $ takes a soliton profile $\varphi_S$ and the Dirac continuum eigenfunctions are called ``soliton states," denoted by $\Psi_E$. Also there is an isolated normalized state at zero energy, $\Psi_0$.
\begin{equation}
\begin{array}{cccc}
H(\varphi_S) \Psi_E &=& E \Psi_E & E \gtrless0\\[.5ex]
H(\varphi_S) \Psi_0 &=& 0
\label{eq2}
\end{array}
\end{equation}

We wish to compute the charge density in the presence of the soliton. This is defined relative to the charge density in the vacuum, where the background field is topologically trivial.
\begin{equation}
\rho ({\bf r}) = \int\limits^{0}_{- \propto}\, d E \bigg(\Psi^\ast_{E}\, ({\bf r})\, \Psi\, (r)  - \psi^\ast_{E}\, ({\bf r})\, \psi_{E}\, (r)\bigg)
\label{eq3}
\end{equation}
The further argument proceeds in its simplest form if the Dirac Hamiltonian possesses an energy reflection symmetry; {\it viz.} if there exists a unitary matrix $R$, which anti-commutes with the Dirac Hamiltonian $HR + R H = 0$.  Then $R$ acting on negative energy states produces positive energy states and vice-versa, while the zero energy state is an eigenstate of $R$.
\begin{equation}
\begin{array}{llll}
R\, \Psi_{\pm E} = \Psi_{\mp E} , &R \, \psi_{\pm E} &=& \psi_{\mp E}\\[.5ex]
&R\, \Psi_0 &=& \pm \Psi_0
\label{eq4}
\end{array}
\end{equation}
(The effective Dirac Hamiltonians for polyacetelyne and graphene possess this property.) As a consequence the charge density at energy $E$ is an even function of $E$, both in the vacuum and soliton sectors: $ \psi^\ast_E \, \psi _E = \psi^\ast_{-E} \, \psi_{-E}, \, \Psi^\ast_E\, \Psi_E = \Psi^\ast_{-E}\, \Psi_{-E},  \Rightarrow \rho_E = \rho_{- E}$; and \eqref{eq3} may be presented as
\begin{equation}
\rho = \frac{1}{2}\ \int\limits^\propto_{-\propto} \, d E \ \left(\Psi^\ast_E\, \Psi_E - \psi^\ast_E\, \psi_E\right).
\label{eq5}
\end{equation}
The vacuum continuum wave functions $\psi_E$ in \eqref{eq5} are complete; the solitonic vave functions $\Psi_E$ are one short of being  complete because the zero energy state $\Psi_0$ is absent. Therefore
\begin{equation}
\rho  \, ({\bf r}) = -\frac{1}{2}\ \Psi^\ast_0 \, ({\bf r}) \, \Psi_0  \, ({\bf r}) 
\label{eq6}
\end{equation}
and the charge $Q$ is
\begin{equation}
Q = \int d {\bf r} \, \rho \, ({\bf r}) = -\frac{1}{2}\, .
\end{equation}

Although this simple argument does not show it, one can prove that the fraction is an eigenvalue without fluctuations, rather than an expectation value with fluctuations --- the latter would not be interesting.

One may easily contemplate Hamiltonians that do not possess energy reflection. For example, take a Hamiltonian with that property, and append to it a term proportional to the previously described matrix $R$. The new Hamiltonian no longer anti-commutes with $R$, and calculation of the fractional charge becomes much more involved.

Two methods have been developed for dealing with this more difficult situation.  One can take a field theoretic approach and calculate in perturbation theory the induced vacuum current in the presence of a soliton $<\bar{\psi}\, \gamma^\mu\, \psi>$. Here $\bar{\psi}\, \gamma^\mu\, \psi$ is the field theoretic current operator, and $<|>$ signifies the field theoretic vacuum in an external background. This determines  the induced charge density and therefore the charge. [The induced current approach is available only for problems without energy reflection symmetry --- the charge in the symmetric case can be obtained in the limit of vanishing asymmetric effects \cite{golston1981}.] Alternatively one can show that the charge density is related to the spectral asymmetry in the soliton sector \cite{niemi1986}.
\begin{equation}
\rho \, ({\bf r}) = -\frac{1}{2}\  \int\limits^\propto_{-\propto} \, d E\  ({\text{sign}\, E}) \ \Psi^\ast_E ({\bf r})\,  \Psi_E ({\bf r})  -\frac{1}{2}\ \Psi^\ast_0\, ({\bf r}) \, \Psi_0 \, ({\bf r})
\label{eq8}
\end{equation}

The calculated vacuum charge in the absence of energy reflection can become an irrational quantity, approaching $-\frac{1}{2}$ when the strength $\epsilon$ of the term violating energy reflection vanishes (But later I shall describe a more involved scenario.) Note that the spectral asymmetry expression \eqref{eq8} immediately yields $Q = - 1/2$ when energy reflection symmetry is present, because in that case the integral vanishes.

All these are fascinating quantum effects, but it remains a challenge to find them in the functional interval.

Next I shall turn away from elaborate formalism, and give a simple counting argument that establishes the fractionalization effect in one dimension, in the linear polymer of polyacetylene.

\begin{figure}[h] 
   \centering
   \includegraphics[width=4in]{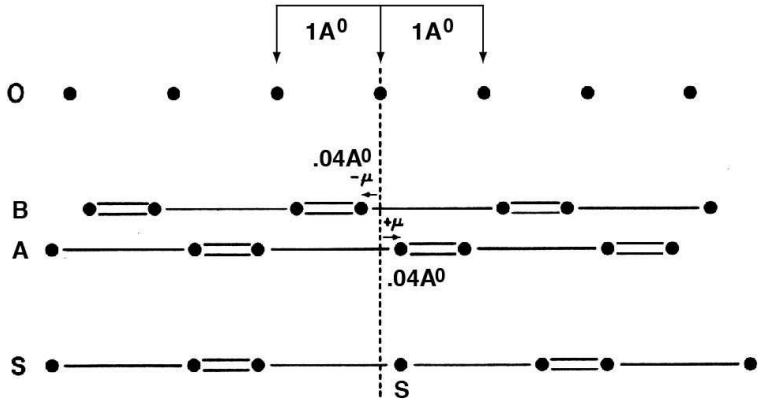} 
   \caption{The equally spaced configuration of carbon atoms in polyacetylene (O) possesses a left-right symmetry, which however is energetically unstable. Rather in the ground states the carbon atoms shift a distance $\mu$ to the left or right, breaking the symmetry and producing two degenerate vacua (A, B). A soliton (S) is a defect in the alteration pattern; it provides a domain wall between configurations (A) and (B).}
   \label{fig1}
\end{figure}

\begin{figure}[ht] 
   \centering
   \includegraphics[width=3in]{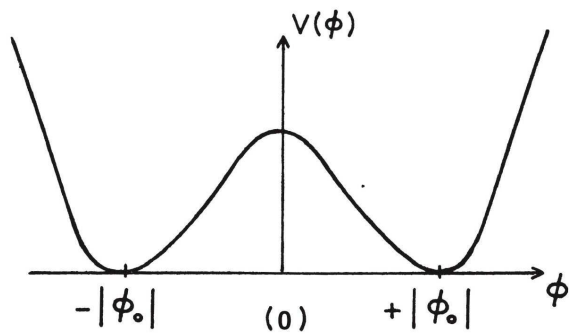}
   \caption{Energy density $V(\phi)$, as a function of a constant phonon field $\phi$. The symmetric stationary point, $\phi = 0$, is  unstable. Stable vacua are at $\phi = + \, |\phi_0|, \text{(A) and}\,  \phi = -|\phi_0|,$ B.}
   \label{fig2}
\end{figure}

To begin a description of polyacetylene, we imagine a rigid array of carbon atoms, about $1 \text{\AA}$ apart, exhibiting a left-right symmetry. One might expect that thermal and quantal fluctuations lead to oscillations about the equidistant equilibrium positions. But in fact something more dramatic happens. The energetics of the system force the atoms to shift by about .04\!\! \AA\  to the left or to the right --- both are allowed due to the left-right symmetry.  This is a consequence of an instability of the rigid lattice, identified by Peierls. Therefore the polyacetylene chain presents two equivalent vacua A and B, (see Fig. \ref{fig1}). 
\eject
\noindent The potential energy of the distortion field (phonon) shows a familiar double well shape: the left-right symmetric point at the origin is unstable; stable configurations at the two minima break the left-right symmetry (see Fig. \ref{fig2}).

\begin{figure}[htbp] 
   \centering
      \includegraphics[width=3in]{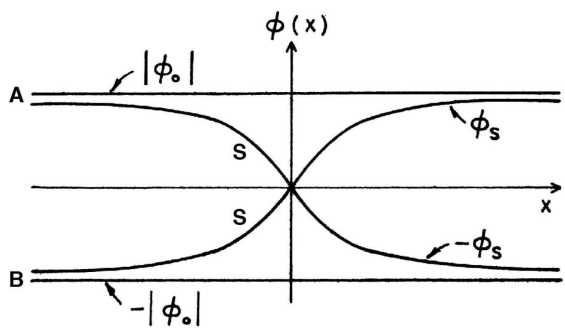} 
   \caption{The two constant fields, $\pm \mid\phi_0 \mid$, correspond to the two vacua (A and B). The two kink fields, $\pm \phi_s$, interpolate between the vacua and represent domain walls.}
   \label{fig3}
\end{figure}

By now it is well known that a double well potential, like in Fig. \ref{fig2}, supports also a kink configuration that interpolates between the two vacua. Physically this represents a defect in the bonding pattern. These profiles are shown in Fig. \ref{fig3}.
\eject

\begin{figure}[ht] 
   \centering
   \includegraphics*[1mm,1mm] [110mm,37mm]{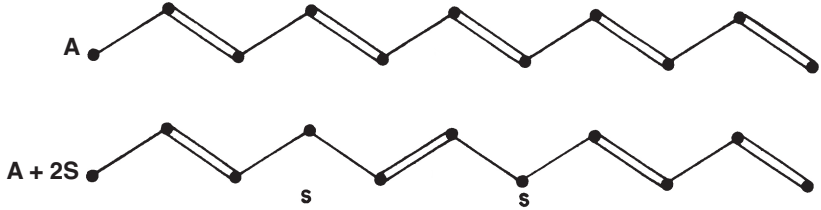} 
   \caption{Two soliton state carries one fewer link relative to the no-soliton vacuum A.}
   \label{fig4}
\end{figure}

Let us now consider a sample of the A configuration, with two solitons --- two defects --- inserted. Let us count the kinks in the A configuration without and with two solitons. These numbers need be considered only in the region between the solitons, because elsewhere the patterns are identical (see Fig. \ref{fig4}).

One observes that in presence of the two solitons, there is one fewer kink. Imagine now separating the solitons to great distance, so that each acts independently. We conclude that the 1-kink deficit must be equally divided, producing a fermion state with number $-\frac{1}{2}$.

In fact this has been indirectly observed. However, the theory has to be elaborated before it confronts experiment. Our argument has ignored spin. Since electrons have spin $\pm \frac{1}{2}$ and two fit in each level, all our results are doubled. The charge defect in the presence of a soliton is $-1$, but there is no net spin since all other electrons are pairwise aligned. Inserting one electron erases the defect, but produces a spin $\frac{1}{2}$ excitation with zero charge. This charge-spin separation, which has been observed experimentally, gives a physical realization to the phenomenon of charge fractionalization in polyacetylene \cite{wpsu}.

The Dirac Hamiltonian matrix, relevant to this 1-dimensional problem, is a $2\times 2$ matrix acting on a 2-component ``spinor", thereby producing a matrix equation in one spatial dimension
\[
H(\varphi) = \alpha  p + \beta \varphi
\]
The kinetic term $\alpha p \, (\alpha \equiv \sigma^3, p \equiv -i \, \partial_x)$ comes from linearizing a quadratic dispersion law around the two points, called ``Dirac points," where it intersects the Fermi surface, thereby giving rise to a 2-component structure. The interaction with the phonon field $ \beta \varphi\ ({\beta \equiv \sigma^2})$ produces a gap with homogenous $\varphi$ and a zero mode with a kink profile for $\varphi$.  The $\sigma^1$ matrix anti-commutes with $H(\varphi)$ and acts as the $R$ matrix that implements the energy reflection symmetry, which is lost when $\epsilon \sigma^1$ is added to the Hamiltonian, thereby producing an irrational charge.

More recently there appeared a 2-dimensional material, for which a similar analysis has been performed. This is graphene, which is described by a hexagonal lattice of carbon atoms that is presented as a superposition of two triangular sublattices see Fig. 5.
\begin{figure}[ht] 
   \centering
\includegraphics[]{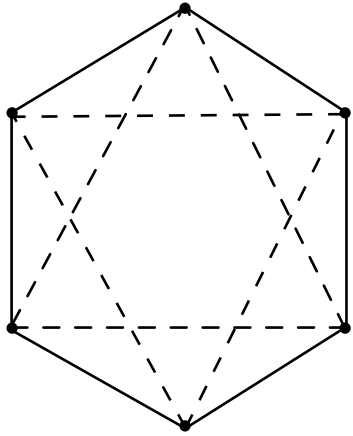}
 \caption{The hexagonal graphene lattice, taken as a superposition of two triangular sublattices.}
   \label{fig5}
\end{figure}
Each sublattice supports two Dirac points, hence a 4-dimensional matrix equation on the two dimensional plane becomes relevant. The background field is taken to be a combination of  scalar $\varphi$ and vector ${\bf A}$  \cite{Hou:2006qc}. The Hamiltonian governing electron motion reads
\[
\begin{array}{ll}
& H(\varphi, {\bf A}) = {\boldsymbol \alpha} \cdot [{\bf p} - \gamma_5 {\bf A}] + \beta [\varphi^r - i\, \gamma_5\, \varphi^i].\\[1.5ex]
\text{Here}& {\boldsymbol \alpha} = \left(\begin{matrix}
							       {\boldsymbol \sigma} & 0 \\
							       0 & - {\boldsymbol \sigma }
							\end{matrix}\right), \
		{\boldsymbol \sigma } = (\sigma^1, \sigma^2); \ \gamma_5 =
							\left(\begin{matrix}
							I & 0\\
							0 & -I
							\end{matrix}\right)\\[3ex]
		&\beta = 				\left(\begin{matrix}
							0 & I\\
							I & 0
							\end{matrix}\right), \ 
		{\bf p} = \frac{1}{i} \, (\partial_x , \partial_y)
							
\end{array}
\]
where $\varphi^r , \varphi^i$ form the real and imaginary parts of a complex scalar field $\varphi = \varphi^r + i \varphi^i$. The matrix that effects energy reflection is $R = \alpha^3 = \left(\begin{matrix}  \sigma^3 & 0 \\ 0 & -\sigma^3 \end{matrix}\right)$. 

The topologically trivial background consist of constant $\varphi$ and vanishing ${\bf A}$. This gives rise to a gap in the energy spectrum. For the topologically non-trivial back ground we take for $\varphi$ and ${\bf A}$  a vortex profile. A zero mode ensues, and charge becomes $-\frac{1}{2}$ (with a single vortex). The mid-gap state is bound with just a vortex configuration for $\varphi$ and vanishing $\bf A$. $\bf A$ is not needed for binding the zero energy state. Its presence does affect the profile of the zero-energy wave function, but not the vanishing of the eigenvalue. However, a vector potential is needed to give the vortex finite energy: the scalar field and vector field vortex configurations separately carry infinite energy, which is rendered finite in the combination. In other words a pure $\varphi$ vortex cannot ``move" because it is infinitely heavy. Evidently this infinity is screened away by the contribution from $\bf A$.

When $ \varepsilon R = \varepsilon \alpha^3 $ is appended to $H$, the energy reflection symmetry is lost, and the induced charge structure becomes interestingly complicated. For  a pure $\varphi$ vortex, without a vector potential, the charge becomes irrational --- as expected and as it happens in the 1-dimensional case. However, including a vector potential in its vortex configuration changes the charge back to $-\frac{1}{2}$ --- its value in the energy reflection symmetric case. Just as for energy, the vector potential screens away a contribution from the scalar field.

The above 2-dimensional story is very interesting in its elegant intricacies. But it must be stated that its validity as a theoretical description of actually realized physical graphene remains to be established. Also as yet there is no experimental verification of the charge fractionalization phenomenon in graphene. Nevertheless it would be most interesting to develop a functional integral analysis of this phenomenon. 

\doefunds

\end{document}